\documentclass[utf8]{FrontiersinHarvard} 

\usepackage{url,hyperref,lineno,microtype,subcaption}
\usepackage[onehalfspacing]{setspace}

\usepackage{graphicx} % Required for inserting images

\def\firstAuthorLast{McIntosh \& Leamon}
\def\Authors{Scott W.\ McIntosh\,$^{1,2,*}$, Robert J. Leamon\,$^{1,3,4}$}

\def\Address{
  $^{1}${Lynker Space, 338 East Market St. Suite 100, Leesburg, VA 20176 USA} \\
  $^{2}${National Center for Atmospheric Research, P.O. Box 3000, Boulder, CO~80307, USA} \\
  $^{3}${University of Maryland--Baltimore County, Goddard Planetary Heliophysics Institute, Baltimore, MD 21250, USA} \\
  $^{4}${NASA Goddard Space Flight Center, Code 672, Greenbelt, MD 20771, USA}
  }

\def\corrAuthor{Scott W. McIntosh}
\def\corrEmail{smcintosh@lynker.com}

\newcommand{\degree}{\ensuremath{^\circ}}

\newcommand{\apj}{    {\it Astrophys. J.}}
\newcommand{\apjl}{   {\it Astrophys. J. Lett.}}

\newcommand{\grl}{    {\it Geophys. Res. Lett.}}

\newcommand{\jgr}{    {\it J. Geophys. Res.}}

\newcommand{\nat}{    {\it Nature}}

\newcommand{\solphys}{{\it Solar Phys.}}
 
\newcommand{\ssr}{    {\it Space Sci. Rev.}} 
\chardef\us=`\_

\newcommand{\pref}{\protect\ref}

\begin{document}
\onecolumn
\firstpage{1}

\title[Connecting Solar `Weather' \& `Climate']{Deciphering Solar Magnetic Activity: Some (Unpopular) Thoughts On the Coupling of the Sun's ``Weather'' and ``Climate''}

\author[\firstAuthorLast ]{\Authors} %This field will be automatically populated
\address{} %This field will be automatically populated
\correspondance{} %This field will be automatically populated

\extraAuth{}% If there are more than 1 corresponding author, comment this line and uncomment the next one.
%\extraAuth{corresponding Author2 \\ Laboratory X2, Institute X2, Department X2, Organization X2, Street X2, City X2 , State XX2 (only USA, Canada and Australia), Zip Code2, X2 Country X2, email2@uni2.edu}

\maketitle

\begin{abstract}
The Sun exhibits episodic surges of magnetic activity across a range of temporal and spatial scales, the most prominent of which is the 11-ish year modulation of sunspot production. Beside the $\sim$170 (min to max) decadal variation in sunspot production there is a less-explored quasi-annual variation in the range of 25-50 sunspots/year in magnitude. In addition, there is there is a slower, $\sim$80 year period, 10-50 variation in the sunspot number, that is commonly referred to as the ‘Gleissberg Cycle.’ Using a suite of contemporary and historical observations we will illustrate these elements of our star’s episodic behavior and present a hypothesis that may provide a consistent physical link between the observed ‘climatic’, ‘decadal’ and ‘seasonal’ magnetic variation of our star.
\end{abstract}

\section{Introduction}

Solar activity is a dynamic and multifaceted phenomenon that strongly influences aspects of the solar system and also of our planet. Indeed, the solar system {\em is} the extended coronal envelope of our Sun. The study of solar activity is essential for understanding space weather, predicting its impacts on Earth, and advancing our knowledge of the Sun's role in the cosmos. Ongoing research and advancements in solar observation technologies continue to deepen our understanding of these intricate processes and their broader implications. 

The Sun’s activity is driven by complex processes occurring within its interior---often hidden from direct observation. These processes manifest in the multi-scalar magnetism that extrudes through the Sun’s optical surface---the photosphere. The solar dynamo operates in the Sun’s convective zone (deep) under the optical surface, where hot plasma rises towards the Sun's surface, cools, and then sinks back into the interior. This convective motion, coupled with the Sun's rotation, creates a complex and turbulent flow of plasma \citep[see, e.g.,][]{Charbonneau2020}. It is then inferred that the Sun's differential rotation, with different latitudinal regions rotating at different speeds, further contributes to the generation of a strong and intricate magnetic field \citep[see, e.g.,][]{Babcock1961,Leighton1969}. 

The principal indicator of solar activity (and the multi-scale magnetism of the solar dynamo) is the ‘solar cycle’ or ‘sunspot cycle’, which lasts approximately 11 years \citep[see, e.g., ][]{Hathaway2010}. The solar cycle is characterized by the rise and fall in occurrence of magnetized sunspots. Similarly, but on a shorter timescale sunspots, active regions, and the solar flares and coronal mass ejections (CMEs) that they generate, exhibit a quasi-annual variation that is most amplified at the peak of the decadal sunspot cycle \citep[see, e.g.,][]{McIntosh2015}. So, in essence the solar cycle, or the process generating it, is the `carrier’ for the secondary, lower-amplitude, oscillation. 

Finally, the Sun and its magnetic spots appear to undergo a multi-decadal ‘oscillation’ in number [see, \citep[see, e.g.,][]{Gleissberg1966, McCracken2000, McCracken2001}. This longer period cycle, the eponymous ‘Gleissberg’ cycle, is most easily observed in the progression of solar (sunspot) maxima over the years and {\em may\/} itself be a carrier for the decadal scale oscillation. In the following sections of the paper we will explore (some of the) manifestation of these ‘cycles’ and a possible explanation for how they may co-exist---effectively connecting the magnetic climate and weather of our star.

This paper does not explore the root causes of the Sun’s dynamo process, instead we will refer the interested reader to some of the excellent reviews on the topic provided earlier \citep{Hathaway2010, Charbonneau2020}. In the section that immediately follows we will present a series of observations that exhibit ‘oscillations’ or ‘modes’ of fluctuation in the Sun’s magnetic activity across temporal scales before discussing a hypothesis that (possibly) links the observations to an underlying phenomena. That underlying phenomena possibly places constraints to the understanding of the Sun’s dynamo process.

\begin{figure}[ht]
\centering
\includegraphics[width=0.75\linewidth]{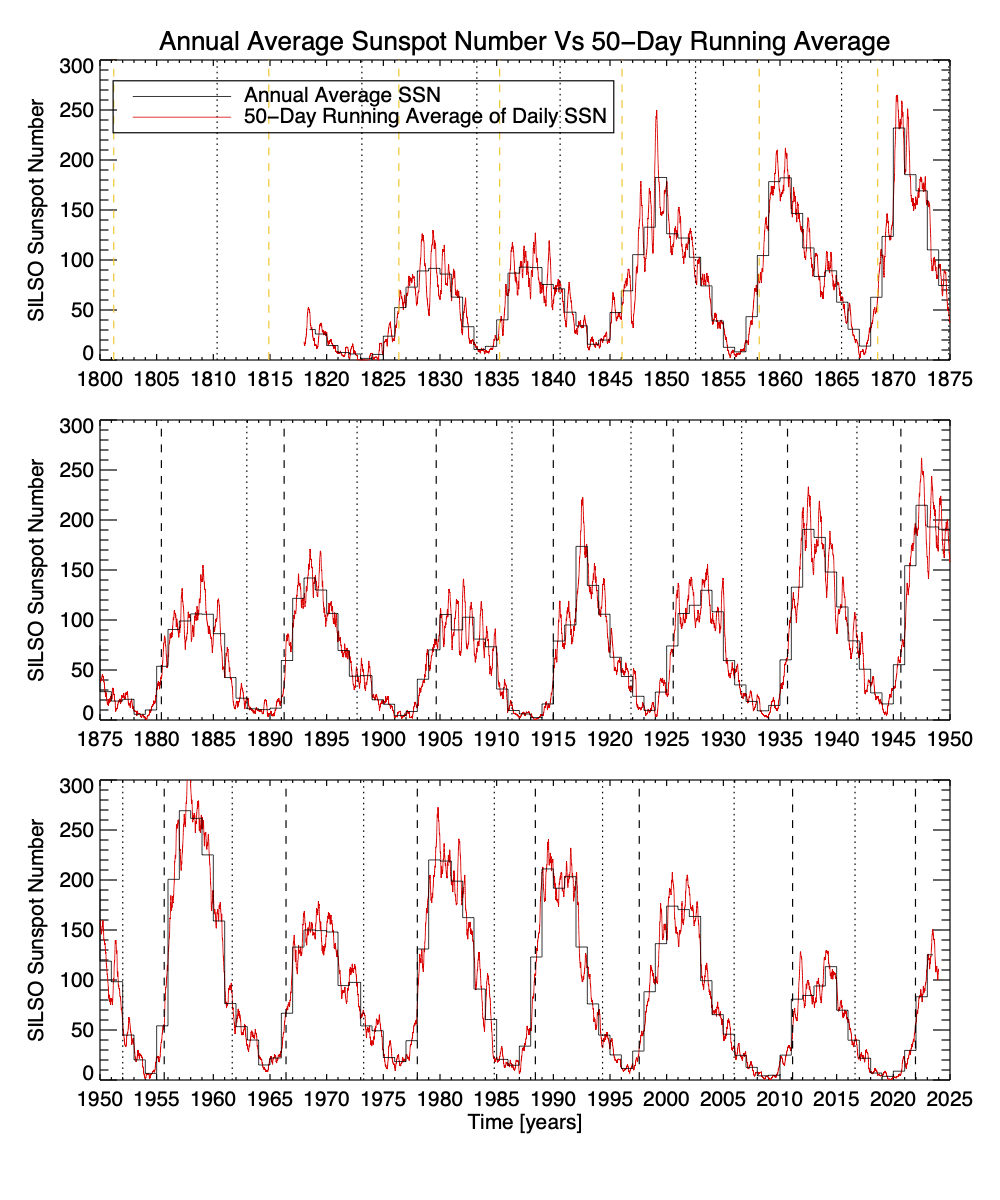}
\caption{The World Data Center (WDC) SILSO v2 sunspot number from 1818 to the present \citep{Clette2015}. Broken into three 75-year epochs we show the annual average total sunspot number (black) and the daily total sunspot number with a 50-day running average applied (red). Black vertical lines are shown to illustrate the occurrence of Hale cycle termination (dashed) and ‘pre-termination’ (dotted) events---see text for further details. Source: WDC-SILSO, Royal Observatory of Belgium, Brussels.}
\label{fig:f1}
\end{figure}

\section{Observational Background \& Motivation}
Figure~\pref{fig:f1} shows the World Data Center (WDC) Solar Influences Data Analysis Center (SILSO) daily total sunspot number (v2) from 1818 to the present \citep{Clette2015}. We contrast the annual averages of the total sunspot number in black with a 50-day running average applied to the daily sunspot number, shown in red. Each of these traces highlights a different mode of solar activity. The first is the prominent (much-studied) modulation of the solar or sunspot cycle where consecutive maxima and minima are separated by approximately 11 years. Further, we see that the amplitude of this mode is not fixed, fluctuating from a minimum value of about 100 to a maximum value of around 250. These last two features are essential characteristics of the sunspot cycle, an 11\-ish year cycle period with an non-negligible oscillating amplitude and have motivated more than a century and a half of academic investigation since first being noticed \citep{Schwabe1844}. 

The 50-day running average of the daily total sunspot number illustrates the presence of a different quasi-annual oscillation in sunspot production of non-negligible amplitude \citep{McIntosh2015}. At times near sunspot maximum this secondary oscillation, sitting on top of the sunspot cycle ‘carrier’ mode, exhibits peak-to-trough variations of order 10-50 sunspots which reduces in amplitude as we approach sunspot minimum, but remains visible.

\begin{figure}[ht]
\centering
\includegraphics[width=1.00\linewidth]{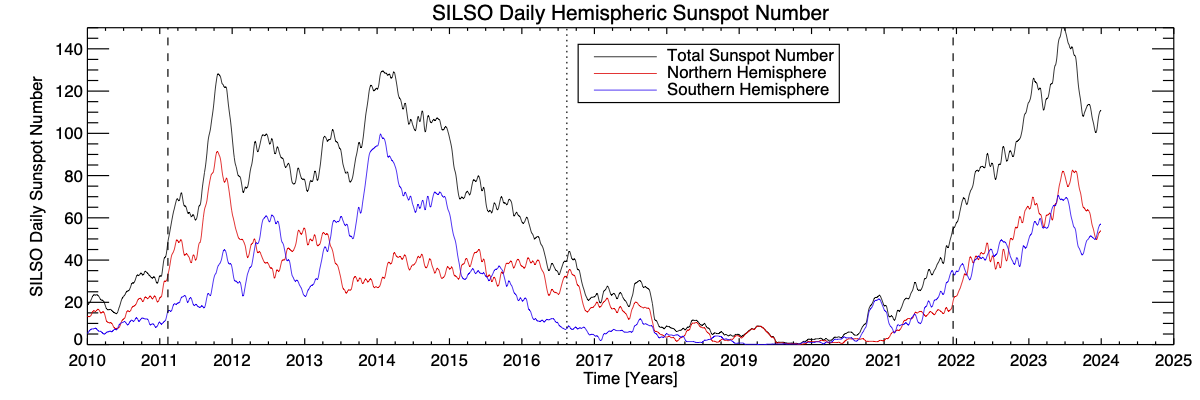}
\caption{The World Data Center (WDC) SILSO v2 hemispheric sunspot number from 2010 to the present. We show the total sunspot number (black), the northern (red), and southern (blue) with a 50-day running average applied---contrast with the red data shown in Fig.~\pref{fig:f1}. Black vertical lines are shown to illustrate the occurrence of Hale cycle termination (dashed) and ‘pre-termination’ (dotted) events---see text for further details. Source: WDC-SILSO, Royal Observatory of Belgium, Brussels.}
\label{fig:f2}
\end{figure}

Zooming in to the timeframe encompassing Sunspot Cycle 24 and the early years of Sunspot Cycle 25 in Figure~\pref{fig:f2} we show the WDC/SILSO 50-day running average applied to the daily hemispheric and total sunspot numbers. While the total sunspot number is shown in black, the northern and southern hemispheric sunspot numbers are shown in red and blue, respectively. In this figure we see that each solar hemisphere exhibits the significant secondary oscillation in sunspot production discussed above. Obviously, the total sunspot number is the cumulation of these two modes and the interested reader can readily compare the black trace in this figure with the red trace of Fig.~\pref{fig:f1}. We see that the oscillations in each hemisphere are not always in phase and that Sunspot Cycle 24 has a very marked hemispheric asymmetry with the northern hemisphere’s sunspot count peaking in late 2011 while its southern counterpart doesn’t peak until more than two years later.

The primary feature of Figure~\pref{fig:f2}, as in Fig.~\pref{fig:f1}, is that the short duration surges in sunspot production can occur at almost any time, even deep into what would be called the solar (or sunspot) minimum phase. For example, the surge in 2017 that gave rise to the largest flares of sunspot cycle 24 \citep[e.g.,][]{2024SoPh..299...39H}. Similarly, there was also a significant amplitude southern hemispheric surge that peaked in late 2020. That final surge preceded the end of the underlying magnetic Hale Cycle, and the triggering of Sunspot Cycle 25 growth by a year. In other words, these significant hemispheric surges in sunspot production appear to be related to, but not necessarily driven by, the process that produces the sunspot number. \citet{McIntosh2015} noted the very strong correspondence between these quasi-annual surges of magnetic flux emergence and the most energetic of solar flares and CMEs in the epoch studied.

\begin{figure}[ht]
\centering
\includegraphics[width=0.65\linewidth]{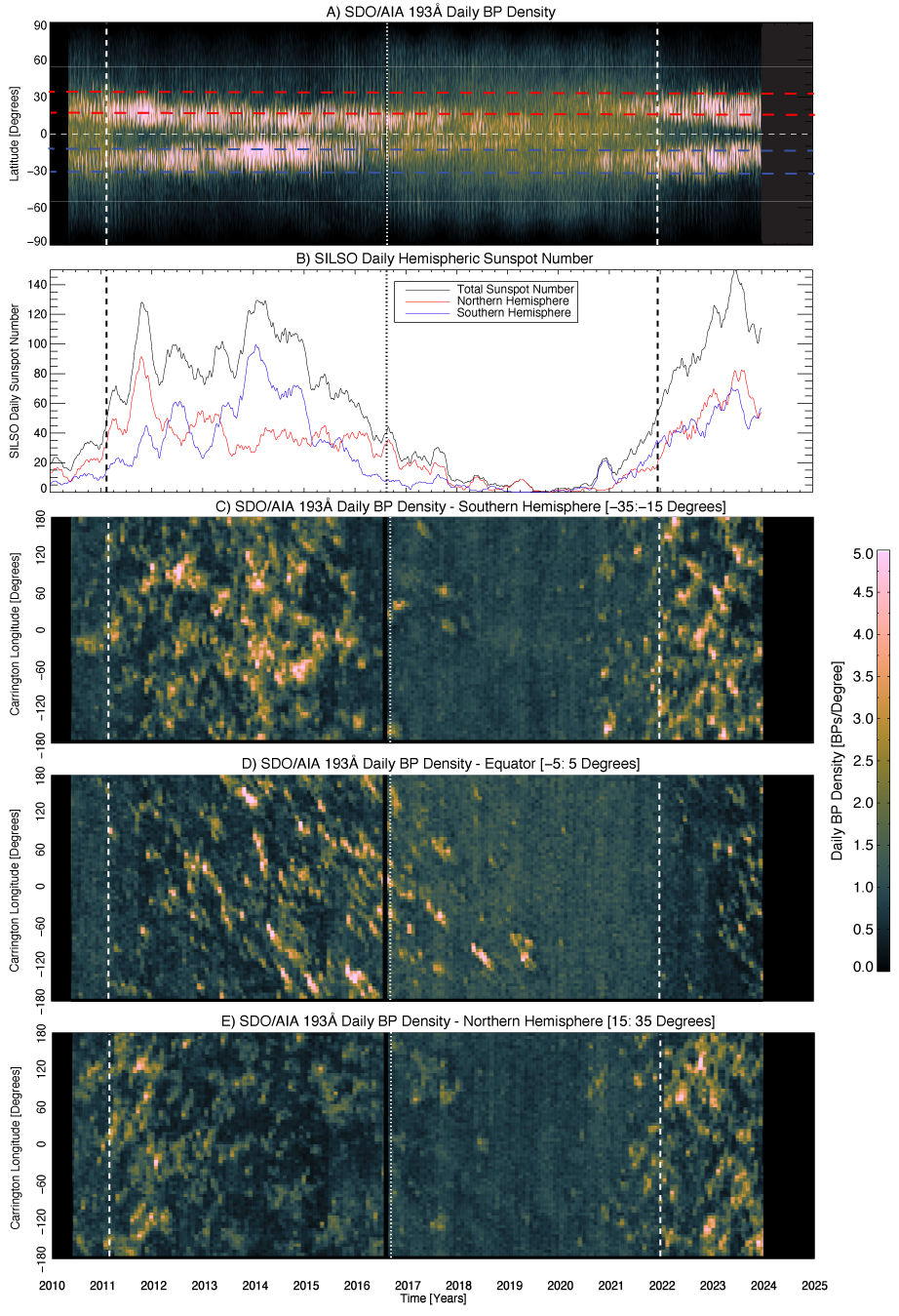}
\caption{Comparing latitudinal and longitudinal evolution of EUV Brightpoint density over the duration of the Solar Dynamics Observatory mission (2010--Present) with the hemispheric and total sunspot numbers over the same epoch. Panel A shows the latitude-time evolution of EUV Brightpoint density, also shown are horizontal lines to designate the range of the northern hemispheric activity band (red: 15\degree:35\degree) and northern hemispheric activity band (blue: -35\degree:-15\degree). Panels C-D show the longitude-time evolution of EUV Brightpoint density sampled over a fixed range of latitudes (‘Hovm\"{o}ller’ diagrams---northern activity band, equatorial band and the southern hemispheric activity band). White vertical lines are shown in each panel to illustrate the occurrence of Hale cycle termination (dashed) and ‘pre-termination’ (dotted) events---see text for further details.}
\label{fig:f3}
\end{figure}

Figure~\pref{fig:f3} begins our exploration into the origins of these magnetic flux surges. We contrast the evolution of the hemispheric sunspot numbers with a marker of meso-scale magnetic flux emergence we have used extensively in the past; extreme-ultraviolet bright points \citep[hereafter BP, e.g.,][]{Mac14,McIntosh2014a}. Further, we consider BP evolution in latitude and longitude, the latter through the utilization of longitude versus time plots over a narrow range of latitudes, or `Hovm\"{o}ller' diagrams \citep{Hovmoller1949}. Panel B of the figure repeats the analysis of Figure~\pref{fig:f2}. Panel A shows the BP density distribution over time, versus latitude, sampled between $\pm$ 5\degree{} of the central meridian from pole to pole. On this figure we draw vertical lines to illustrate landmark features in the epoch, dotted horizontal lines at $\pm$ 55\degree latitude, a dashed line at the solar equator, dashed red lines (15\degree{} and 35\degree) and dashed blue lines (-35\degree{} and -15\degree) to delineate activity zones that will be used to construct the Hovm\"{o}ller diagrams in panels C through E. The equatorial region we define to be $\pm$ 5\degree{} about the equator. Each slice in a Hovm\"{o}ller diagram (in this case oriented horizontally with time running left to right) is the average BP density in an annulus, or ring, of over the defined range of latitudes. 

Figure~\pref{fig:f3} is constructed for the reader to scan vertically and allow cross-comparison of the meso-scale magnetic flux emergence patterns in latitude and longitude with time. In addition, we observe the subtle changes in the background magnetism that are outlined by vertical lines on the figure, and those that precede it---these lines are relevant to the picture that will develop. First, we can look at surges of flux emergence in the northern hemisphere (like that in late 2011) which is spread across seven longitudes (for four to six rotations) with that in the southern hemisphere which was much more concentrated (80-100\degree longitude). Similarly, the early 2014 peak in southern hemispheric activity is observed in latitude and longitude. One can also see the diagonally-oriented stripes that represent ‘trains’ of slowly migrating magnetic flux that have been associated with the presence of (magnetized) Rossby waves \citep[e.g.,][]{McIntosh2017, Dikpati2017, Dikpati2020}. 

\begin{figure}[ht]
\centering
\includegraphics[width=1.00\linewidth]{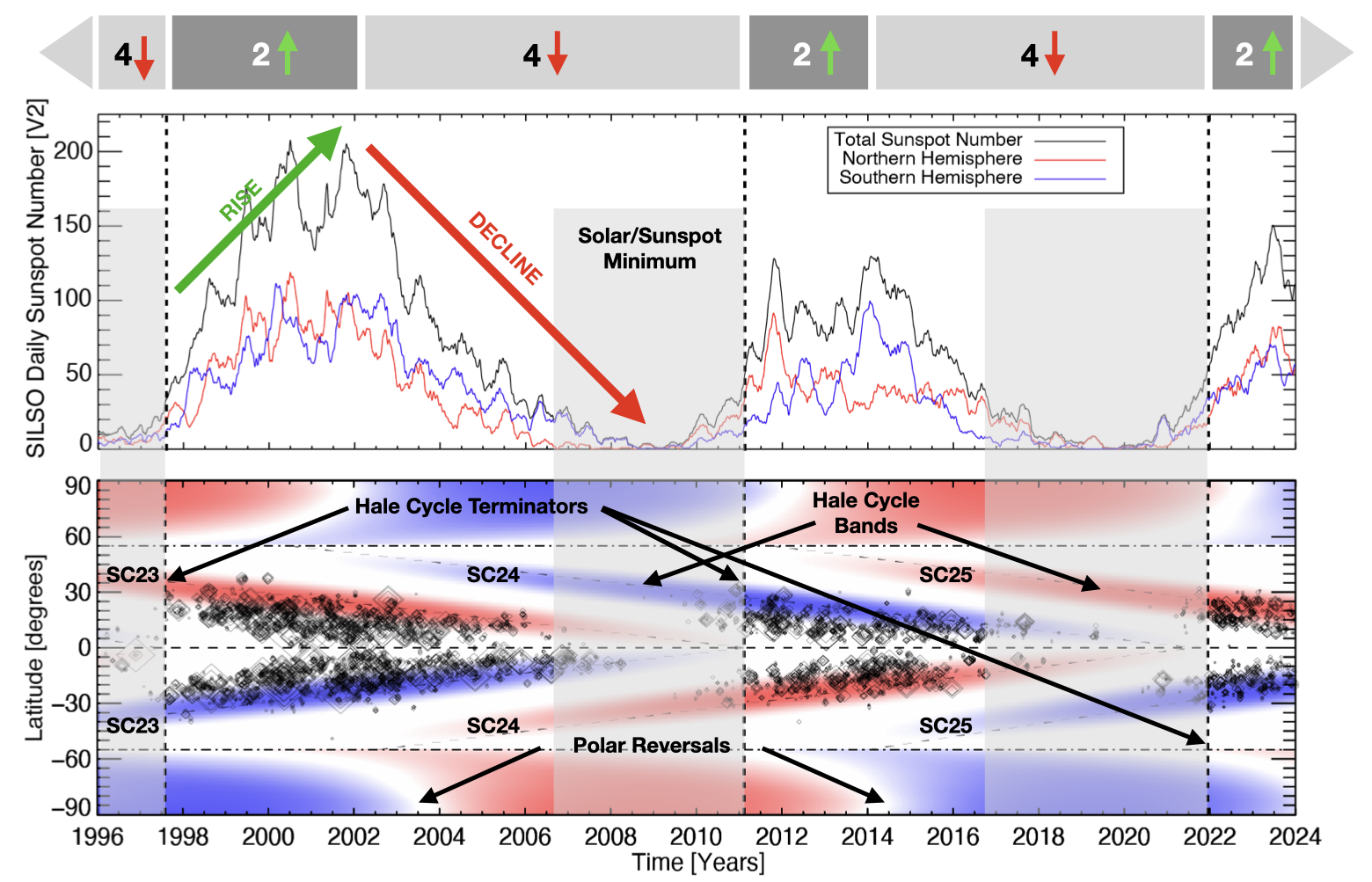}
\caption{An infographic of the past two sunspot cycles to illustrate the landmarks of the Sunspot Cycle and how they tie to the fundamental Hale Cycle, including the canonical landmarks of sunspot/solar maximum and solar/sunspot minimum. The lower panel shows the sunspot “butterfly” diagrams superposed on top of the (data-inferred) magnetically polarized Hale Cycle bands. There are four sets of Hale cycle bands visible in this time frame: those hosting sunspot cycles 22 through 25.}
\label{fig:f4}
\end{figure}

Finally, another phenomena is observed in the panels of Figs.~\pref{fig:f1} to~\pref{fig:f3} and we use the (data-driven) infographic in Fig.~\pref{fig:f4} to explain its occurrence and the possible role it plays in helping shape the decadal scale variation we observe as the Sunspot Cycle. Figure~\pref{fig:f4} illustrates the landmarks of the Sunspot Cycle and how they tie to the fundamental underlying 22-year magnetic Hale Cycle \citep{McIntosh2022}, including the canonical landmarks of sunspot/solar maximum and solar/sunspot minimum. 

The lower panel shows the sunspot ``butterfly’’ diagram superposed on top of the (data-inferred) magnetically polarized Hale Cycle bands \citep{Mac14, McIntosh2022} with the expansion of the hemispheric and total sunspot numbers shown in Fig.~\pref{fig:f2} to include Sunspot Cycle 23. We see that there are four sets of Hale cycle bands visible in this time frame: those hosting Sunspot Cycles 22 through 25. With this illustration it is straightforward to see and consider the overlap in latitude and time of the Hale Cycle bands. 

After starting their long (17--19 years) journey to the equator from $\sim$55\degree{} latitude at the time of (hemispheric) solar sunspot maxima, the magnetic bands of the Hale Cycle coexist within a hemisphere and across the equator for a long period of time. We call the reader’s attention to times of sunspot minima (1994--1997; 2007--2011; 2017--2022) when there are clearly four oppositely polarized Hale cycle bands within 40\degree{} of the Sun’s equator---and the Sun makes very very few spots during these epochs. We interpret this observation in the following way: the intra- and extra-hemispheric (magnetic) connections at this time result in strong interactions between the bands with insufficient free magnetic flux latitudinally (but not necessarily longitudinally, see above) to enable the buoyant rise of coherent large scale magnetic flux features \citep{Mac14}. This state continues until the two, oppositely polarized. bands nearest the equator end abruptly in the (Hale Cycle) ‘termination’ event \citep[][, or ``terminator’’]{Mac14,McIntosh2019,McIntosh2022}. The resulting rapid change (drop) in BP density at the equator (at all solar longitudes) can be observed in panels A and C of Figure~\pref{fig:f3}. The terminator events for each Hale Cycle are shown in the figures of this section as dashed vertical lines (1997, 2011, and 2021). 

The termination of the Hale Cycle bands at the equator leaves the two remaining, oppositely polarized, bands at mid-latitudes (between 30\degree{} and 35\degree{} latitude). We observe an instant (within a solar rotation) reaction and observe the very rapid growth of BPs and sunspots, this event marks the start of the ``ascending’’ phase for the sunspot cycle. At the time of the terminator, in addition to the rapid growth of sunspots at mid latitudes, we observe the start of the polar reversal process at $\sim$55\degree{} latitude, commencing the ($\sim$2-3 year) march of the high latitude neutral delineation of the polar region new Hale Cycle from the previous one as easily visualized through the polar crown filament’s motions \citep{McIntosh2021,Leamon2022}. The two-band, post-terminator, ascending phase state below 55\degree{} latitude continues until around the time of the solar polar magnetic field reversal and the solar maximum that often happens concurrently \citep[see, e.g.,][]{1991ApJ...375..761W, 2023SSRv..219...31Y}. 

Note that, after the polar reversal epoch, an extra set of oppositely polarized magnetic bands is starting to emerge at around 55\degree{} latitude. These new bands belong to the next Hale Cycle and are beginning their long (17--19 year) march to the equator and we are reaching half way through the Hale Cycle progression. Again, as a result of the new intra-hemispheric (magnetic) interactions being introduced, the amount of free magnetic flux is reduced in each hemisphere and the number of spots being produced on the lower latitude bands begins to drop. The frequency of sunspot production continues in a slow decay that eventually results in sunspot minimum epoch some 4--6 years later. During this epoch, it has been observed that the progression of the magnetized bands towards the equator progressively slows \citep[see, e.g.,][]{Hathaway2010, 2016ASPC..504..241T, 2020Ge&Ae..59.1016T}.

\citet{Leamon2022} \citep[and][]{2020GeoRL..4786524C, 2020GeoRL..4787795C, 2021ApJ...917...54C} explored the start of the quiescent phase of the sunspot cycle using superposed epoch analyses and found that it occurs around the time that the higher latitude Hale Cycle bands cross 45\degree{} latitude with almost complete reduction in sunspot production by the time it has progressed to 40\degree{} latitude. The start of the quiescent phase, the ‘pre-terminator,’ for each Hale Cycle are shown in the figures of this section as dotted vertical lines \citep{Leamon2022}. This progression continues until the four-band state is ended by the next terminator event at the equator and the magnetic polarity pattern of the Hale Cycle bands is back to where we began. 

Basically, we use this conceptual infographic to deduce that we {\em can} observe sunspots in a locations where there is significant imbalanced magnetic flux relative to the background conditions---where longitudinal and longitudinal effects can contribute to the degree of imbalance.

We stress that it is, as yet, unknown what might be `special' about 55\degree{} latitude on the Sun. We observe that the magnetic bands of the Hale Cycle repeatedly appear to grow and then bifurcate there \citep{McIntosh2021}. Whatever the process, the nature of that latitude seems to be fundamental to the evolution of Hale and Sunspot Cycles alike, but why this progression happens is not entirely clear! This is a topic we will return to below. Similarly, we do not fully understand the process by which the Hale Cycle bands can move faster or slower in during this evolution, noting that conventional dynamo models invoke meridional circulation changes to explain this observation \citep[e.g.,][]{Charbonneau2020}.

\subsection{Rossby Waves Revisited}
Rossby waves are a type of planetary wave that plays a significant role in large-scale atmospheric and oceanic circulation \citep{Rossby1939}. They are large-scale waves that typically form in the Earth's atmosphere or oceans due to the Coriolis effect as a consequence of the planet's rotation. Rossby waves are characterized by their meandering, wavelike patterns that can extend horizontally across large distances.

Terrestrial (atmospheric) Rossby waves play a crucial role in the movement of weather systems and the development of long-term weather patterns. They contribute to the development of highs and lows, influencing the movement of air masses and the formation of weather fronts. Rossby waves are closely related to the formation and behavior of jet streams in the Earth's atmosphere (high-altitude, fast-flowing air currents known as jet streams). Like the terrestrial atmosphere, there are also global-scale oceanic Rossby waves. These waves are slower and have larger wavelengths than their atmospheric counterparts and play a role in redistributing heat in the oceans and hence affecting ocean currents.

Understanding Rossby waves is essential for meteorologists and oceanographers as they significantly impact the Earth's climate and weather patterns. Changes in the behavior of (atmospheric) Rossby waves can have profound effects on regional weather events, such as heatwaves, droughts, and storms, making them an important aspect of climate research and prediction. Similarly, following \citet{McIntosh2017} and developed further by \citet{Dikpati+McIntosh2020}, the Sun’s interior exhibits similar rotationally-driven waves, but they are strongly modified by toroidal magnetic fields in the solar tachocline \citep{Dikpati2020}. 

These waves influence the process of magnetic flux emergence \citep{Dikpati2017}, including the quasi-annual surges observed in the daily sunspot numbers (Figs.,~\pref{fig:f1}--\pref{fig:f3}) and the resulting (significant) eruptive event occurrence \citep{McIntosh2015}. Therefore, monitoring and characterizing these waves through the observed patterns, using objects like BPs, can be used to anticipate when and where magnetic flux may emerge and events may occur on a rotational timescale and beyond.

\section{A Connecting Hypothesis?}
In the preceding paragraphs we have discussed a range of phenomena connected to the production and modulation of (magnetic) sunspots that, we hypothesize, result from the magnetic systems that comprise the Sun’s Hale Cycle. Conventional understanding places the emergence of sunspots (and hence active regions) as a consequence of the buoyancy of the larger-scale magnetic system exceeding that of the gas (plasma) in which it is embedded \citep[see, e.g.,][]{2009LRSP....6....4F,2014LRSP...11....3C}. 

Our objective is to reconcile the inferred long-range (latitudinal) interactions of the Hale Cycle bands that give rise to events like the terminator, pre-terminator, and shape the sunspot cycle, with the (longitudinal) occurrence of Rossby wave driven flux emergence. Our hypothesis extends the straightforward local picture of buoyant magnetic flux emergence to one where the global-scale ‘plasma’ pressure balance (a spatio-temporal battle between magnetic pressure and gas pressure) throughout the Sun’s interior is {\em the\/} determining factor on where magnetic flux emerges. Basically, where there is balance in the total plasma pressure, in longitude and/or latitude, the Sun cannot produce sunspots and conversely, you can have conditions where there is imbalance in longitude and/or latitude and the Sun can produce spots at those locations.

From a latitudinal perspective, the repeated occurrence of the terminator and pre-terminator events highlight their significance and interpretative origin. Further, these events appear to have a critical role in keeping time in, or synchronizing, the system. As a result they impact the shape by modulating the increase and decrease of sunspot production on the Hale Cycle magnetic flux systems \citep[see, e.g.,][and the work that follows]{Mac14}. While the precise physics of the Hale Cycle terminator events remains unresolved it is hard to argue with the impact and/or synchronization of the event. Twenty four of them have been observed in sunspot data, fourteen have been observed in imaging data, and four have been observed in magnetography data. Initial exploration of the terminator events \citep[see, e.g.,][]{Dikpati2019}, and, by extension, the means by which the flux systems `teleconnect’ are encouraging and consistent with the observations. Initial numerical investigations into the paradigm of `band interaction’ are instructive in shaping future self-consistent investigations \citep[see, e.g.,][]{Belucz2023}. In a longitudinal sense, we have clearly demonstrated that there are (magnetized) Rossby waves present. Those waves intrinsically affect the (local) pressure balance and hence, the reason that there are persistent ‘active’ longitudes of flux that slowly meander and travel pro-grade relative to (differential) rotation \citep[see, e.g.,][]{Dikpati2017, Dikpati2019, Dikpati2020}. 

Both of these activity ‘modes’ exist in the system concurrently. We readily observe in Fig.~\pref{fig:f1}, over many cycles, that the shorter period Rossby waves are almost omnipresent while the longer period ‘decadal’ mode  due to the inferred interaction of the Hale Cycle bands has not, apparently, stopped over the epoch of human observation.

\section{Discussion}
The presence of events like the terminator (and also the pre-terminator) are indicative of a ‘magnetic teleconnection’ process taking place---a synchronization of magnetic flux emergence globally \citep[e.g.,][]{McIntosh2020}. These rapid global changes are a hard condition to replicate in contemporary, diffusive, dynamo models \citep[e.g.,][]{Charbonneau2020}. From this we infer that the global-scale magnetic flux systems of the Hale Cycle may be stronger than the few kilogauss values generated in those same conventional models \citep[see, e.g.,][]{Belucz2023}. Further, that these fields are not necessarily passively trapped in, and/or following, the large scale circulatory patterns of the solar interior as per conventional dynamo models. 

It unclear at what magnitude these magnetized systems would have to be to exhibit the timescales of global change observed and/of the flow-field feedbacks that may be induced by such strong fields. Instead, we infer that they are of a magnitude sufficient to interact over distance and hence experience attractive and repulsive forces which could also govern their migratory motion. This, by extension, would appear to suggest a picture of the Sun’s interior where the ‘thin flux tube’ and ‘frozen-in’ conditions \citep[e.g.,][]{Charbonneau2020} are not uniformly applicable to global scale phenomena---i.e, there are locations in longitude and latitude where the conditions are not applicable and other approximations and/or methods should be used to replicate their essential characteristics. 

We wonder if the global scale magnetic fields present in the Hale Cycle may be sufficient to actually impact the apparent migratory speed of those bands as a result of their multi-range interactions. In the community literature many column inches of emphasis have been placed on the apparent migratory speed of the sunspot butterflies to the equator with the steepness of the ascending/descending phases of the sunspot cycle \citep{Hathaway2010}. Through Fig.~\pref{fig:f4} we know that the number and proximity of the interactions happening in the solar interior is constantly changing and the confirmations have intrinsically different complexity and strength in ascending and descending phases.

We also infer that these global-scale magnetic systems, if they can affect the migratory motions, may also be strong enough to self-consistently create convective imbalances that modify differential rotation anomalies observed as the `torsional oscillation’ \citep[see, e.g.,][]{1980ApJ...239L..33H}. We note that, in conventional models, significant and rapid changes of the meridional circulation speed have been invoked to explain the change in migratory speeds observed \citep{2011Natur.471...80N}, but the authors question the degree to which this process is physically possible in a massive rotating body (see later).

The following subsections explore two (curious, but well known) examples to add weight to our underlying hypothesis that magnetic field related pressure imbalances shape and modulate solar activity across timescales. 

\begin{figure}[ht]
\centering
\includegraphics[width=1.00\linewidth]{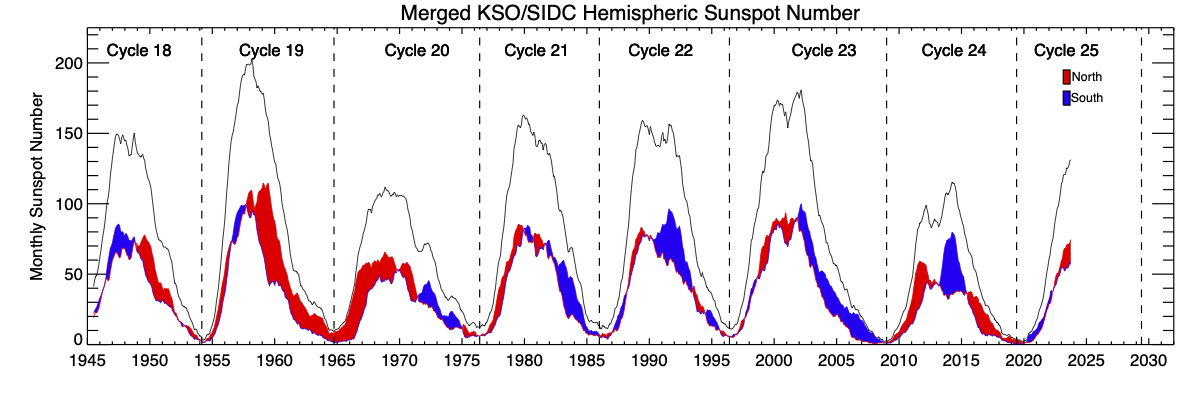}
\caption{The World Data Center (WDC) SILSO v2 monthly hemispheric sunspot number from 1946 to the present \citep{Clette2015}. The variation in the total (black), northern (red) and southern (blue) hemispheric sunspot numbers are shown and the difference between the hemispheric sunspot numbers is colorized red or blue to illustrate which hemisphere, north or south, respectively has an excess.}
\label{fig:f5}
\end{figure}

\subsection{The Gleissberg Cycle}
The Gleissberg cycle is the name given to a long-term variation in the solar activity cycle \citep{1944TeMAE..49..243G}. It is characterized by periodic fluctuations in the number of sunspots over a time span of approximately 80 to 90 years and is superimposed on the sunspot cycle \citep[see, e.g.,][and others]{gleissberg1958eighty, 2003JGRA..108.1003P}. A published characteristic of the Gleissberg cycle is that is amplitude is typically smaller than that of the 11-year solar cycle and represents significant variability in solar activity over longer timescales.

\begin{figure}[ht]
\centering
\includegraphics[width=1.00\linewidth]{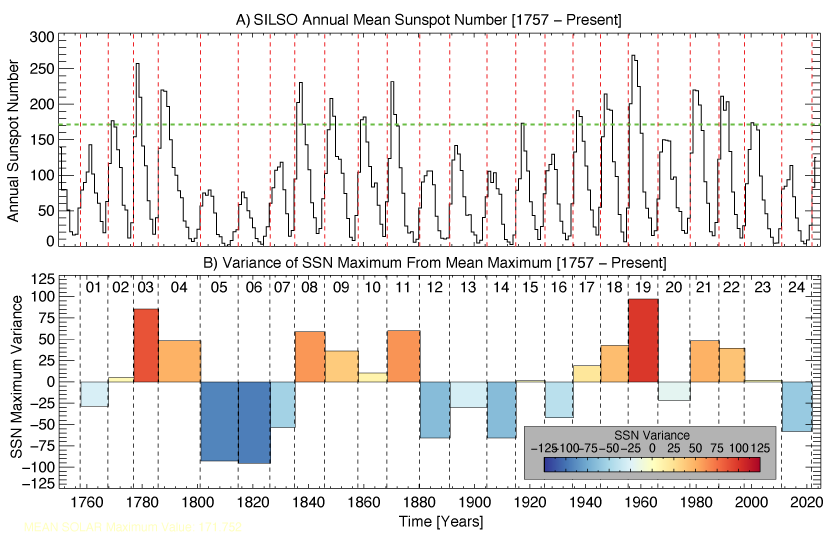}
\caption{The World Data Center (WDC) SILSO v2 monthly total sunspot number from 1749 to the present \citep{Clette2015} is shown in the Panel A. The horizontal green dashed line signifies the mean sunspot maximum value over this epoch 172. This mean value is then used in Panel B, a block diagram to illustrate the deviation of the sunspot cycle maxima from the mean. In each panel the vertical dashed lines illustrate the Hale Cycle terminator events.}
\label{fig:f6}
\end{figure}

Figures~\pref{fig:f5} and~~\pref{fig:f6} show two possible representations of the Gleissberg cycle in the historical sunspot record. The first, in Fig.~\pref{fig:f5}, shows the modulation of hemispheric sunspot number `leader-follower' relationship---i.e., when solar hemispheres exhibit systematic imbalance in production of sunspots \citep[e.g.,][]{2013ApJ...765..146M}. The colorization of the difference between the northern and southern hemispheric sunspot number into red or blue signifies which hemisphere presently has more sunspots, north or south respectively. Unfortunately, the published hemispheric sunspot number is not of sufficient duration to unambiguously illustrate a number of 80\-ish year Gleissberg cycles, but it is obvious these sunspot cycles exhibit long-term, multi-cycle, systematically north-south imbalances.

The second, possibly more straightforward, visualization of the Gleissberg cycle comes from the historical sunspot records from 1749 to the present shown in Fig.~\pref{fig:f6}. Panel B of the figure shows a block diagram to illustrate the deviation of the sunspot cycle maxima from the mean of all maxima (172, shown by the green dashed line in Panel A). On this much longer temporal baseline one can see an approximately 80-90 year period oscillation in the sunspot maxima. Other, longer geophysical data records exhibit similar periodicities \citep[eee, e.g.,][]{2003JGRA..108.1003P}

While the exact mechanism driving the Gleissberg cycle is not well understood it is clearly influenced by the dynamics of the solar dynamo. How do the variations in the strength and organization of the solar magnetic field may contribute to the observed long-term changes in sunspot activity? In the context of this manuscript we consider a mode of variation where we are experiencing the manifestation of a slow `sloshing' in the imbalance of solar magnetism across the equator.

% The Gleissberg cycle, first identified in 1862, strengthens and weakens the 11-year cycle over the course of a century (shown in yellow). One paper posits that the Gleissberg pattern is caused by a slow swaying of the sun's magnetic pole.

\subsection{Tail Wags Dog?: The Curious Case Of Sunspot Cycles 5 \& 24}
Another (curious) topic relates to the low (significantly below average) amplitude of Sunspot Cycle 24 \citep[e.g.,][]{2020JSWSC..10...60P}, the unusually long sunspot minimum between Sunspot Cycles 23 and 24 \citep[e.g.,][]{2010ASPC..428.....C, 2012GeoRL..39.4102L}, and the start of the (early) nineteenth century's Dalton Minimum and what happened between Sunspot Cycles 4 and 5 \citep[e.g.,][]{2015JSWSC...5A..30O, 2021NatSR..11.5482M}. How and why are these topics related?

\begin{figure}[ht]
\centering
\includegraphics[width=0.7\linewidth]{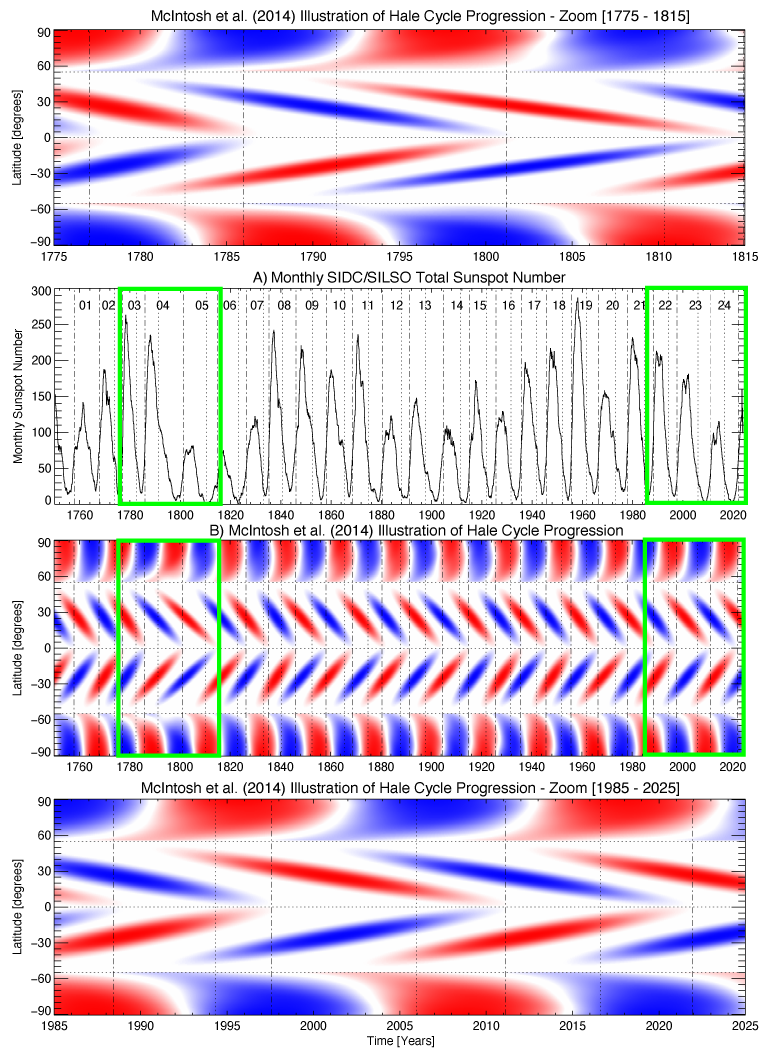}
\caption{The World Data Center (WDC) SILSO v2 monthly total sunspot number from 1749 to the present \citep{Clette2015} is shown in the Panel A. Panel B shows the data-driven schematic of the underlying magnetic Hale Cycle for the same epoch \citep{Mac14}. Shown on these two panels are green boxes, one for the 1775--1815 epoch and another for the 1985--2025 epoch. The uppermost and lowermost panels of the figure show zoomed-in versions of the schematic for those two epochs. Each panel uses vertical dashed lines to indicate the Hale Cycle termination events. In addition we show vertical dotted lines to indicate the 0.6 value of the \citet{Leamon2022} unit Hale Cycle.}
\label{fig:f7}
\end{figure}

We ask the reader to consider Fig.~\pref{fig:f7}. Panel A shows the Monthly sunspot number from 1749 to the present (see Fig.~\pref{fig:f6}) and Panel B shows the corresponding data-driven schematic of the underlying magnetic Hale Cycle (see Fig.~\pref{fig:f4}). From these schematics, derived from the times of the termination events and the sunspot maxima \citep{Mac14}, we zoom in the 1775--1815 epoch (top panel) and 1985--2025 epoch (bottom panel). We see from the collection of schematics that the terminator events that end sunspot cycle 4 and sunspot cycle 23 are well separated in time from those preceding them. As per \citet{McIntosh2019} the longer terminator separations imply that the upcoming cycle strength would be reduced. The original inference being that the slide of a terminator event to the right `ate into the growth time of the upcoming cycle' \citep{Mac14}. Incidentally, the terminator separations times leading to sunspot cycle 5 (15.217 years) and 24 (12.825 years) are two of the three longest in the 25 events studied \citep{McIntosh2019} and definitely represent extrema.

So, the terminator events slide to the right. This has an impact on the upcoming sunspot cycle amplitude: reducing it. The question is, is there a cause producing this effect? Or, is the significant reduction in upcoming cycle amplitude the result of an event (or events) that occurred during the sunspot cycle? One that destroyed magnetic flux and slowed down the progression of the Hale cycle? In the language we have used above, did something happen that significantly reduced the amount of available magnetic field in Sunspot Cycles 4 and 23? 

\begin{figure}[ht]
\centering
\includegraphics[width=1.0\linewidth]{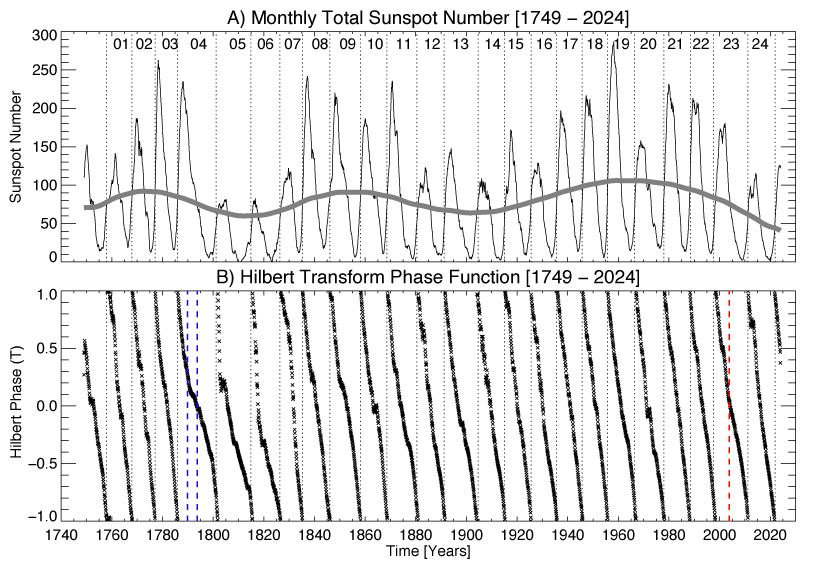}
\caption{The World Data Center (WDC) SILSO v2 monthly total sunspot number from 1749 to the present \citep{Clette2015} is shown in the Panel A. The thick gray curve in Panel A shows the 40-month RLOWESS trend for the sunspot data. Panel B shows the Hilbert transform of the trend-removed phase function for the same epoch \citep{McIntosh2019}. The vertical dotted lines in the panels indicate the Hale Cycle termination events. There three vertical colored lines, one in 1789, 1793/4 (both blue) and another in 2003 (red).}
\label{fig:f8}
\end{figure}

Consider Fig.~\pref{fig:f8} where we contrast the monthly total sunspot number and the corresponding Hilbert transform phase function \citep[cf. Fig.~2 of][]{McIntosh2019}. \citet{Leamon2022} already suggested that the knee visible in the Hilbert phase function in late 2003 (dashed vertical red line) was the result of the Halloween storms \citep[e.g.,][]{2004EOSTr..85..105L, TSURUTANI20061583}. Indeed it is acknowledged in the literature that a significant amount of magnetic flux was reduced in the sequence of events occurring in October and November 2003 \citep[see, e.g.,][]{2005ApJ...623L..53M, 2022LRSP...19....2C} in addition to the observation of \citet{2011SoPh..272..337T}. Did the Halloween storms of 2003 result in the reduction of amplitude of Sunspot Cycle 24?

Perhaps more interesting are the events of the late eighteenth century. Indeed, Sunspot Cycles 4/5 received significant attention as the Sun entered into the Dalton Minimum with Sunspot Cycle 4 possessing such an odd, extended shape, where inferences made that there was a missing sunspot cycle \citep[e.g.,][]{2015JSWSC...5A..30O}. \citep{McCracken2001} identified three significant events occurred in the late 1780s and early 1790s, illustrated by blue dashed vertical lines in Fig.~\pref{fig:f8}B. The storm that reached the Earth on November 13th into the 14th of 1789 produced aurorae that were readily observed between 30 and 17 degrees geographic latitude \citep{2021AdSpR..68.2320R}. The events of 1793 and 1794 also produced significant low latitude aurorae \citep{2005SoPh..231..157V}, the first of which produced twice the signal in the ice core record than the second, approximately the same as the 1789 event \citep{McCracken2001}.

While the relative solar and geomagnetic signature of these events versus that of 2003 is unclear, it is clear that there these events must have been driven by significant magnetic energy on the Sun. The extended and odd shape of Sunspot Cycle 4, the resulting kneed Hilbert phase function and the significant reduction of Sunspot Cycle 5 would appear to be consistent with the conditions that impacted Sunspot Cycle 24. Interestingly, \citet{McCracken2001} identified a series of significant events in their ice core samples that, while not as closely packed as those in the 1780/90s persisted through the 1820s while few were detected between 1760--1780 or between 1830--1850 where sunspot numbers are at or slightly above average. Speculation about whether or not these processes contribute to the phenomenon of Grand Minima, or at least the Dalton Minimum, far exceeds the scope of this manuscript, but may be a subject of future investigation \-- similarly, with the uniqueness of the McCracken's interpretation of the ice cores \citep[see, e.g.,][]{Mekhaldi2017}.

\section{Conclusion}
We have discussed a range of events that would appear to infer that the global-scale magnetic fields of the sun's interior interact with one another. This interaction shapes the manifestation of surface solar activity that we see and feel as the sunspot cycle on a decadal scale through the magnetic Hale Cycle. The relentless sloshing of magnetic imbalance between the solar hemispheres will subtly impact the progression and interaction of the Hale Cycle systems and the resulting sunspot cycles to affect the climatic scale of solar variability. 
The bands of the Hale Cycle exhibit buoyant `planetary' waves which result in epochs of enhanced magnetic activity on top of the decadal scale variability. These epochs give rise to the most intense episodes of eruptive space weather. In response to the most of extreme solar magnetic weather events, the process by which the Hale Cycle progresses appears to slow down, sometimes significantly. This slowing of the hale Cycle would appear to impact upcoming cycle strengths. So, while we have a situation where numerous modes of variability are present, there seems also to be processes which feedback from the shorter timescales into the longer ones. 

Taken in concert, these observations would appear to support a general thesis that the global-scale magnetic fields of the Sun's interior interact strongly across spatial scales (intra- and extra-hemispherically) to shape variability across temporal scales. However unpopular this thesis may be, the concept of two overlapping Hale Cycles, their interplay, and their termination at the equator as the key, fiducial, time of solar activity is simultaneously consistent with the greatest number of disparate observations of variable solar output.

%% To paraphrase Sherlock Holmes [Conan Doyle, 1890]
%% ``How often have I said to you that when you have eliminated the impossible, whatever remains, however unpopular [improbable], must be the truth?''

\section*{Acknowledgements}
SMC's work at NCAR was supported by the National Center for Atmospheric Research, which is a major facility sponsored by the National Science Foundation under Cooperative Agreement No. 1852977. RJL acknowledges support from NASA's Living With a Star Program.  Sunspot data from the NOAA Space Weather Prediction Center and the World Data Center SILSO, Royal Observatory of Belgium, Brussels. We thank Frank Howell [[K4FMH] for his support, encouragement and enduring wisdom.

\section*{Conflict of Interest Statement}
The authors declare that the research was conducted in the absence of any commercial or financial relationships that could be construed as a potential conflict of interest.

\section*{Author Contributions}
All authors conceived the experiment, analyzed the results and reviewed the manuscript.

\section*{Funding}
Details of all funding sources should be provided, including grant numbers if applicable. Please ensure to add all necessary funding information, as after publication this is no longer possible.

\section*{Acknowledgments}
This is a short text to acknowledge the contributions of specific colleagues, institutions, or agencies that aided the efforts of the authors.

% \section*{Data Availability Statement}
% The datasets [GENERATED/ANALYZED] for this study can be found in the [NAME OF REPOSITORY] [LINK].
% Please see the availability of data guidelines for more information, at https://www.frontiersin.org/about/author-guidelines#AvailabilityofData

\bibliographystyle{Frontiers-Harvard} 
%\bibliography{sample}

\end{document}